\documentclass{JFM-FLM_Au}

\usepackage{graphicx}
\usepackage{subfigure}
\usepackage{amsmath}
\usepackage{amssymb}
\usepackage{listings}
\usepackage{array}
\usepackage{xcolor}

\usepackage{subfigure}
\usepackage{amssymb}

\lefttitle{}
\righttitle{Journal of Fluid Mechanics}

\title{On the Reynolds analogy for high-speed rough-wall flows: implications for wall modelling}

\author{Michele Cogo\aff{1,3}, Davide Depieri\aff{2}, Matteo Bernardini\aff{1} \and Francesco Picano\aff{2,3}}

 \affiliation{\aff{1}Department of Mechanical and Aerospace Engineering, Sapienza University of Rome, via Eudossiana 18, 00184 Rome, Italy
 \aff{2}CISAS "Giuseppe Colombo", Università degli Studi di Padova, via Venezia 1, 35131, Padova, Italy
 \aff{3}Department of Industrial Engineering, Università degli Studi di Padova,
via Venezia 1, 35131, Padova, Italy
}

 \corresau{Michele Cogo, \email{michele.cogo@unipd.it}}

\begin{document}
\maketitle

\begin{abstract}
We study the validity of the generalized Reynolds analogy (GRA) in compressible turbulent boundary layers over prism-shaped roughness by mining direct numerical simulation data of Mach 2 and Mach 4 compressible turbulent boundary layers with adiabatic and cooled surfaces. Although the direct influence of roughness strongly disrupts the near-wall coupling between momentum and energy, we show that this breakdown is confined to the roughness sublayer. 
Above this layer, the enthalpy and velocity fields recover a smooth-wall-like similarity, and the GRA becomes asymptotically valid by naturally accounting for roughness-enhanced wall shear stress and heat flux. 
Building on these results, we propose a GRA-based wall model for predicting heat transfer over rough surfaces, which is coupled with a drag-predictive physics-based method developed for prism-shaped roughness by means of compressibility transformations.
\end{abstract}

 \vspace{-1cm}

\section{Introduction}
The interaction between compressibility, surface roughness, and heat transfer poses a longstanding challenge in the understanding and modelling of wall-bounded high-speed flows. These effects are central to many high-speed applications and directly impact drag and heat transfer, which are critical to the structural integrity of vehicles operating in and out of the atmosphere.
In particular, the accurate prediction of thermal loads
is inherently tied to our ability to assess the relation between velocity and temperature fields, which are nonlinearly coupled through density. 
However, surface roughness further complicates this coupling, and our predictive capabilities for such flows still rely heavily on semi-empirical correlations that do not generalize across the vast parameter space spanned by Mach number, wall-thermal conditions, and roughness geometry.

In smooth-wall compressible boundary layers, mean thermodynamic quantities are related to the mean momentum field by the Reynolds analogy, independently deduced by Crocco and Busemann \citep{crocco1932sulla,busemann1931gasdynamik} under the assumption of unity Prandtl number $Pr$.
This relation provides a link between the mean total enthalpy $\bar{H}$ and the mean streamwise velocity $\bar{u}$, relating temperature and velocity by means of a simple quadratic function.
This relation was later extended by \citet{van1951turbulent} for turbulent boundary layers, and several subsequent works incorporated important realistic physical effects, such as deviation of $Pr$ from unity, diabatic conditions and more complex surface effects \citep{walz1969boundary,duan2011direct}.

A comprehensive extension of the classical analogy is provided by the generalized Reynolds analogy (GRA) proposed by \citet{zhang2014generalized}. The GRA introduces a general recovery enthalpy $H_g$ and an effective turbulent Prandtl number $Pr_e$, which is assumed to be approximately unity throughout the boundary layer and largely insensitive to Mach number or wall thermal conditions. This framework yields a universal enthalpy--velocity relation for smooth-wall compressible turbulence, and its validity has since been assessed across a broad set of DNS databases, including channel flows, pipe flows, and boundary layers with both adiabatic and cooled walls \citep{cogo2023assessment}.

From an engineering standpoint, one of the most important observations of the extensive work dedicated to the Reynolds Analogy is the universality of the Reynolds analogy factor $s=2C_h/C_f$, which is shown to hold approximately constant when coupled with the Prandtl number ($sPr\approx0.8)$, with weak dependence on the Mach number and wall temperature condition \citep{zhang2014generalized}. Here, $C_f=2\tau_w/(\rho_\infty u_\infty^2)$ is the skin friction coefficient and $C_h=q_w/(\rho_{\infty}u_{\infty}c_p(T_w-T_r))$ is the Stanton number, whose relation through $s$ has far-reaching implications for reduced-order modelling, given that it directly relates the wall shear stress $\tau_w$ and heat transfer $q_w$.

In this context, it is well-known that the presence of surface roughness clearly affects the value of $s$ \citep{hill1980measurements,Modesti_2022}, being usually lower than the reference smooth wall case. The general consensus is that the increase in skin friction caused by roughness is greater than the corresponding increase in heat transfer. This is because the additional wall-shear stress is transmitted to the surface as a form-drag on the individual asperities, while the heat flux is only controlled by the thermal conductivity \citep{owen_heat_1963}.
Several studies have been conducted to account for this effect with semiempirical relations \citep{owen_heat_1963,chen_compressible_1972,hill1980measurements}, which nevertheless suffer from poor generalizability.

Notwithstanding the added layer of complexity added by rough surfaces, it is important to note that $s$ can be interpreted as an integral measure of the Reynolds Analogy \citep{wenzel2022influences}, and in this sense its departure from universality for rough wall flows does not directly imply that the analogy between averaged momentum and energy equations is invalid locally everywhere in the flow.
%In turn, the relationship between generalized recovery enthalpy $H_g$ and velocity $U$ provided from the GRA can be locally evaluated throughout the boundary layer.

The present study aims to build on this consideration by assessing the validity of the GRA throughout the entirety of the boundary layer, which has important implications for the prediction of the wall heat transfer on compressible flows over rough walls.
%which is also inherently tied to the existence of an outer-layer similarity for the velocity field with the smooth-wall reference.
To this end, we leverage a novel DNS dataset of compressible turbulent boundary layers over prism-shaped roughness based on the works of \citet{cogo2025a,cogo2025b}, encompassing Mach numbers 2 and 4 and two wall temperature conditions: adiabatic and cold wall.
Building on our findings, we propose a wall model for compressible rough-wall flows by extending the drag-predictive method introduced by \citet{yang2016exponential} for incompressible flows, which explicitly accounts for the sheltering mechanism induced by prism-shaped roughness.

\section{Computational setup}
The present study employs a novel dataset of direct numerical simulations (DNS) of supersonic, zero-pressure-gradient turbulent boundary layers at free-stream Mach numbers \(M_\infty = 2\) and \(M_\infty = 4\), over aligned cubical roughness elements. For each Mach number, simulations are performed under adiabatic \((\partial T/\partial y = 0,\ \Theta \approx 1)\) and cold isothermal wall conditions \((\Theta = 0.25)\), yielding four rough-wall cases, denoted M2A, M2I, M4A, and M4I. Each rough-wall case is accompanied by a smooth-wall reference simulation at the same \(Re_\tau\), generated using the model of \citet{manzoor2024estimating} and listed in Table~\ref{tab DNS parameters}.
All rough-wall cases share the same roughness geometry, consisting of equally spaced cubical elements of height \(k\) with streamwise and spanwise spacing \(2k\), identical to that used in \citet{cogo2025a}. 

The DNS are performed using the open-source solver STREAmS \citep{bernardini2023streams}, which solves the compressible Navier--Stokes equations for a viscous, heat-conducting, calorically perfect gas on a Cartesian grid. Molecular viscosity follows Sutherland’s law with \(T_\infty = 220\,\mathrm{K}\), and thermal conductivity is defined via a constant Prandtl number \(Pr = 0.72\). The roughness geometry is handled using a ghost-point-forcing immersed boundary method, previously validated for similar configurations \citep{cogo2025a,cogo2025b}.

The computational domain consists of a smooth-wall inflow region generating a fully turbulent boundary layer of initial thickness \(\delta_{\mathrm{in}}\) via recycling--rescaling, followed by a rough-wall section starting at \(x = 55\,\delta_{\mathrm{in}}\). All simulations use the same grid resolution of \(20240 \times 556 \times 1408\) points in the streamwise, wall-normal, and spanwise directions, with wall-normal stretching to resolve the roughness sublayer. Further details of the numerical setup are provided in \citet{cogo2025a,cogo2025b}.
Reynolds statistics are obtained by combining spanwise averaging, temporal averaging over at least \(500\,\delta_{\mathrm{in}}/u_\infty\), and averaging over a short streamwise window of four roughness wavelengths. Below the roughness crest, intrinsic (fluid-only) spatial averaging is employed.

\begin{table}
\centering
\begin{tabular}{cccccccccc}
\hline 
\hline
Color & Case & $M_\infty$ & Surface & $\Theta$  & $T_w/T_r$ & $Re_\tau$ & $\delta_{99}/k$ & $k^+$ \rule{0pt}{2.6ex} \rule[-1.2ex]{0pt}{0pt}\\
\hline
\textcolor[RGB]{128, 128, 128}{\raisebox{0.5ex}{\rule{1cm}{2pt}}} & M03 \citep{cogo2025a}  & 0.3 & Rough&1.0 & 1.0 & 1600 & 28.5 & 56\rule{0pt}{2.6ex}\\
\textcolor[RGB]{5, 170, 255}{\raisebox{0.5ex}{\rule{1cm}{2pt}}} & M2A & 2 & Rough& 1.0   &1.0 & 1574 & 27.7 & 57 \rule{0pt}{2.6ex}\\
\textcolor[RGB]{249, 132, 74}{\raisebox{0.5ex}{\rule{1cm}{2pt}}} & M2I  & 2 & Rough& 0.25 & 0.69 & 1588 & 28.9& 55\rule{0pt}{2.6ex}\\
\textcolor[RGB]{39, 125, 162}{\raisebox{0.5ex}{\rule{1cm}{2pt}}}& M4A & 4 & Rough& 1.0   &1.0& 1650 &27.6&60 \rule{0pt}{2.6ex}\\
\textcolor[RGB]{209,0,0}{\raisebox{0.5ex}{\rule{1cm}{2pt}}}& M4I  & 4 & Rough& 0.25 & 0.44& 1508 & 29.9 &51\rule{0pt}{2.6ex}\\
\hline
\hline
\end{tabular}
\caption{Summary of parameters for DNS study at selected stations. Here, $\Theta=(T_w-T_\infty)/(T_r-T_\infty)$ is the diabatic parameter, $T_w/T_r$ the wall-to-recovery temperature ratio, $Re_\tau=\rho_wu_\tau \delta_{99}/\mu_w$ the friction Reynolds number, $k$ the roughness height.}\label{tab DNS parameters}
\end{table}

\vspace{-0.4cm}
\section{Validity of the GRA on rough surfaces and outer-layer similarity}\label{sec:validation}
The formulation of the Generalized Reynolds Analogy (GRA) proposed by \cite{zhang2014generalized} introduces a direct relation between mean velocity $\bar{u}$ and total enthalpy defect $\bar{H}_g - \bar{H}_w$, which reads
\begin{equation}\label{eq zhang enthalpy}
 \bar{H}_g-\bar{H}_w =U_w\bar{u},
\end{equation}
where $H_g=c_p T + r_g u^2/2$ is a general recovery enthalpy with the recovery factor $r_g$, which is an extended version of the classical recovery factor $r$ used by \citet{walz1969boundary}.
Equation \eqref{eq zhang enthalpy} shows that the proportionality between mean enthalpy and velocity fields is modulated by a constant velocity scale $U_w=-Pr \bar{q}_w/ \bar{\tau}_w$.
It can be shown from \citet{zhang2014generalized} that the general recovery factor $r_g$ and $U_w$ are related through the expression
\begin{equation}\label{eq generalized recovery factor}
    r_g=\frac{\bar{T}_w-\bar{T}_\delta}{\bar{u}^2_\delta/(2c_p)}+2\frac{U_w}{\bar{u}_\delta},
\end{equation}
yielding a constant value of $r_g$ across the boundary layer, which naturally reduces to the recovery factor $r$ for an adiabatic wall (where $U_w=0$).

Equation \eqref{eq zhang enthalpy} establishes a local relationship between $\bar{H}_g$ and $\bar{u}$ that is constrained to the wall fluxes $q_w$ and $\tau_w$ through their ratio in $U_w$. This coupling of mean flow variables with wall fluxes is essential for predicting heat transfer over rough walls and is absent in classical relations such as that of \citet{walz1969boundary}. The latter does not depend on heat transfer or wall shear stress, except in the degenerate adiabatic case ($q_w=0$), for which \citet{zhang2014generalized} reduces to \citet{walz1969boundary}.

\begin{figure}%[t]
  %\centerline{\fbox{\vbox to 8pc{\hbox to 10pc{}}}}
  \centering
  \includegraphics[trim={0 0 0 0},scale=0.8]{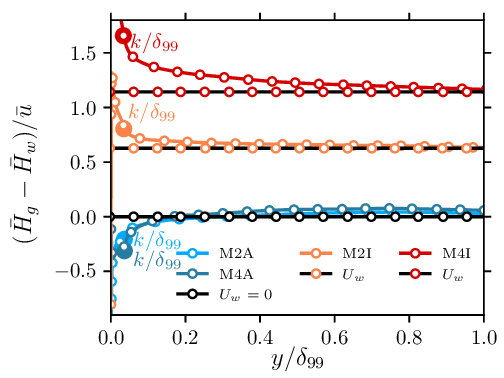}
  \caption{Profiles of the total enthalpy to velocity ratio $(\bar{H}_g-\bar{H}_w)/\bar{u}$ compared to their respective velocity constant scale $U_w$, as per Eq. \eqref{eq zhang enthalpy}, as function of $y/\delta_{99}$
  Here, $U_w=-Pr ( q_w/\tau_w)$, and for adiabatic cases $U_w=0$. The markers indicate the roughness height $k/\delta_{99}$ for each case.}
  \label{H_rg_zhang}
  \vspace{-0.2cm}
\end{figure}

Before elaborating the implications of this point for wall modelling (see Section \S \ref{sec:model}), we focus on the general validity of Eq. \eqref{eq zhang enthalpy}, presented in Figure \ref{H_rg_zhang}. Here, we compare the ratio $(\bar{H}_g-\bar{H}_w)/\bar{u}$ with the predicted value of the velocity constant $U_w$, which is representative of both sides of Eq. \eqref{eq zhang enthalpy} after a simple redistribution of terms.

For all cases, it is apparent that close to the wall, and especially below the roughness crest ($y<k$), the term $(\bar{H}_g-\bar{H}_w)/\bar{u}$ has a clear departure from the constant predicted $U_w$ from the GRA, even for adiabatic cases M2A and M4A where $U_w=0$. This is a region where the GRA is clearly invalid in the presence of roughness, especially if evaluated at the wall as classically done in smooth-wall flows.
However, above the roughness crest the mismatch gradually reduces, and different cases recover their respective $U_w$, even though with different rapidity.
Cases M2A, M4A and M2I recover a good agreement with the expected constant value closer to the wall, whereas case M4I has a slower asymptotic decay.
Nevertheless, this result is of great importance because it shows that beyond the direct influence of roughness there is a region of the flow, the outer layer, where GRA recovers excellent predictive capabilities, establishing a direct relationship between mean flow variables (enthalpy and velocity) and wall fluxes. 

Here, it is key to note that our assessment of the GRA applied to rough-wall flows is local, whereas the vast majority of the studies dedicated to the Reynolds analogy focus on the Reynolds analogy factor $s$, which we argue is more restrictive given that it incorporates the integral behaviour of the boundary layer \citep{wenzel2022influences}.

Conceptually, we can better describe our observation by considering a volume-averaging approach of the NS equations \citep{breugem2005direct}. If the flow is averaged over a small spatial volume that exceeds the roughness scale,  additional forcing terms will appear in both momentum and energy equations accounting for the presence of roughness. 
As thermal and momentum transport are clearly affected in different ways by the presence of roughness, there is a break of the near-wall similarity of momentum and energy that underlies the GRA arguments.
Although roughness introduces forcing terms in the near-wall equations, these decay rapidly when the averaging volume no longer intersects the geometry, and far from the wall the governing equations recover the same differential form as the smooth-wall counterpart.

It is important to note that this argument is valid if roughness effects can only be indirectly felt far from the wall by a change in the effective boundary conditions, under the Townsend outer-layer similarity hypothesis \citep{townsend1980structure}.
For compressible velocity fields, the classical approach to obtain such similarity is the use of compressible transformations, that aim to account for variations of mean flow properties in order to recover the incompressible profiles, for which outer-layer similarity over certain roughness patterns is much more established \citep{cogo2025a}.

In this study, we consider the classical \citet{van1951turbulent} transformation, here denoted by the subscript ‘VD’, which accounts for mean density variations in order to yield an incompressible-like velocity profile $u_{VD}$, such that $u_{VD} = \int_0^{u} \sqrt{\bar{\rho}/\bar{\rho}_w} du$. 
This particular transformation has been selected given its simple formulation, which can be easily incorporated in rough-wall models, and given the fact that only the log layer region is of interest, where a density-based scaling is physically consistent.

\begin{figure}
  %\centerline{\fbox{\vbox to 8pc{\hbox to 10pc{}}}}
  \centering
  \hspace{-2cm}
	\subfigure[][]{\includegraphics[scale=1]{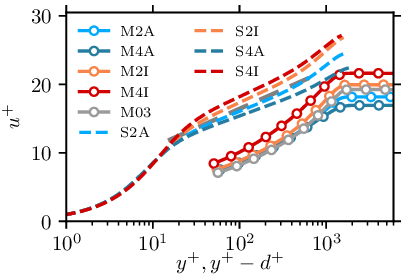}}\qquad
  \hspace{0cm}
	\subfigure[][]{\includegraphics[scale=1]{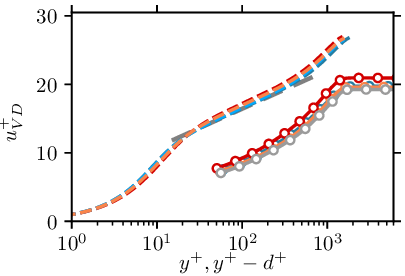}}\qquad
  \hspace{-2cm}
  \vspace{-0.4cm}
  \caption{Comparison between the mean velocity profile $u^+$ as function of $y^+$ before (a) and after (b) applying the transformation of \citet{van1951turbulent}. Rough-wall cases are shifted in the wall-normal direction by the virtual origin $d$. Grey lines represent the log law $u^+=(1/\kappa) \ ln(y^+)+5.2$. The subsonic case M03 from \citet{cogo2025a}, which has the same roughness pattern, is included for reference.}
  \label{up}
\end{figure}

Figure \ref{up} reports both smooth wall profiles (S2A, S2I, S4A, S4I), and rough-wall ones (M2A, M2I, M4A, M4I), in both classical inner units, Panel (a), and by using \citet{van1951turbulent} transformation, Panel (b).
Here, the wall-normal coordinate used for rough-wall profiles is shifted by the same virtual origin $d=k$ \citep{chung2021predicting}, consistent with the previous work of \citet{cogo2025b}.

When observed in classical inner units, Panel (a), both smooth and rough wall profiles are sensitive of the compressibility effects and wall temperature conditions, especially in the log layer. Here, while smooth wall profiles have a well-established incompressible reference, the log law, rough-wall counterparts are compared to the nearly-incompressible case provided by \citep{cogo2025a}, which shares the same roughness pattern and similar flow conditions.
When the \citet{van1951turbulent} transformation is applied to all cases, we observe a good collapse of both smooth and rough profiles with their respective incompressible reference, with only a minor mismatch for the case M4I, a sign that for our setup compressibility effects are still well captured by velocity transformations.
This observation is another important building block for wall modelling efforts, providing a way to leverage the much more extensive theoretical framework existing for incompressible flows by using compressibility transformations.

\section{Formulation of a wall model}\label{sec:model}
In this section, we propose a novel wall model for compressible flows over distributed prism-shaped roughness, capable of predicting the wall shear stress $\tau_w$ and heat flux $q_w$ given local knowledge of the flow dynamics in the outer layer, the roughness geometry and the wall temperature.
%, typical of Wall-Modelled Large-Eddy-Simulations (WMLES).

First, we focus on establishing a connection between the outer-layer validity of the GRA, discussed in Section \S \ref{sec:validation}, and its implications for wall modelling.
By using the definition of $U_w$, and defining a general recovery temperature $T_{rg} = H_g/c_p$, we can rewrite Eq. \eqref{eq zhang enthalpy} as
\begin{equation}
    c_p (\bar{T}_{rg}-\bar{T}_w) = - Pr  \frac{q_w}{\tau_w} \bar{u}.
\end{equation}
By defining the friction temperature as $T_\tau=q_w/c_p\rho_w u_\tau$, and recalling that $\tau_w =\rho_w u_{\tau}^2$, we can further write 
\begin{equation} \label{walz 3}
    \frac{\bar{T}_{rg}-\bar{T}_w}{T_\tau}=-Pr\frac{\bar{u}}{u_\tau}.
\end{equation}
Equation \eqref{walz 3} now better displays how the GRA represents a relation between mean local flow variables, $\bar{T}_{rg}$ and $\bar{u}$, and the expected heat flux and drag, encoded in $T_\tau$ and $u_\tau$.
Recalling the expression of $T_{rg}=\bar{T}+r_g \bar{u}^2/(2 c_p)$, Eq. \eqref{walz 3} can provide an estimate of $T_\tau$ (hence the wall heat flux $q_w$) once mean temperature $\bar{T}$ and the ratio $\bar{u}/u_\tau$ are known at a certain location of the flow.
Most importantly, the ability to infer the wall heat flux $q_w$ solely from its indirect impact on flow variables outside the roughness sublayer is a cornerstone of the present model. Without this capability, rough-wall modelling would require explicitly resolving additional forcing terms in the averaged energy equation induced by unresolved roughness, greatly increasing model complexity and undermining its practical applicability.

Regarding the mean temperature $\bar{T}$, we assume that a parabolic dependency with velocity still exists outside the direct influence of roughness
\begin{equation}\label{eq griffin}
    \bar{T}=b_0+b_1\bar{u}+b_2\frac{\bar{u}^2}{2},
\end{equation}
where the coefficients $b_0,b_1$ and $b_2$ can be determined by evaluating the flow field in known locations. Classically, wall models frequently assume that information about the wall, $\bar{T}=T_w$ and $\bar{u}=0$, and the edge of the boundary layer, $\bar{T}=T_e$ and $\bar{u}=u_e$, are known, and in the context of Wall-Modelled Large-Eddy-Simulations (WMLES) an additional point can be provided by the matching location of choice, $\bar{T}=T_m$ and $\bar{u}=u_m$, where the field is assumed to be adequately resolved.
This assumption is backed by different studies \citep{Modesti_2022,cogo2025b}, which observed the existence of an approximate quadratic relationship between the temperature and velocity in the presence of roughness.

We now turn our attention to the estimation of the velocity ratio $\bar{u}/u_\tau$, equivalent to $u^+$. Recalling the discussion in Section \S \ref{sec:validation}, the additional complexity of having to deal with compressibility effects can be strongly alleviated by applying a velocity transformation, such as the one of \citet{van1951turbulent}. This shifts the problem in finding an estimate of the incompressible velocity profile $u^+_{VD}$, effectively incorporating the compressible and wall temperature effects by accounting for mean density variations.

The Van Driest transformation can be easily incorporated in the differential form of the logarithmic law of the wall for the fully-rough regime \citep{chung2021predicting}, which reads
\begin{equation}\label{differential log law}
    \frac{d\bar{u}}{dy}=\frac{u_{\tau}}{\kappa}\frac{1}{y-d}\sqrt{\frac{\bar{\rho}_w}{\bar{\rho}}}%+\frac{u_\tau}{\delta}\frac{\Pi}{k}\pi\ \sin \bigg(\pi \frac{y}{\delta} \bigg),
\end{equation}
where $\kappa$ is the von Karman constant and $d$ is the virtual origin, which needs to be modeled. 
Here, the density ratio can be evaluated using the equation of state for an ideal gas $p=\rho R T$, where $R$ is the gas constant and $p$ is the pressure, which is assumed to be equal to its value at the edge $p=p_e$ under the zero-pressure-gradient assumption.
The apparent simplicity of Eq. \eqref{differential log law}, which can be integrated in order to find $\bar{u}/u_\tau$, hides all the complexity of rough-wall flows, which is the estimation of the integration constant, different from a roughness pattern to another. 
This is an open problem for rough-wall flows, and the reader is referred to the comprehensive reviews of \citet{jimenez2004turbulent, chung2021predicting} for more details.

We consider the physics-based model of \citet{yang2016exponential}, which offers a solution to this problem for prism-shaped roughness by providing an estimate of the flow velocity at the roughness crest $\bar{u}_k$, which serves as the lower integration bound for solving Eq. \eqref{differential log law}.

The velocity $\bar{u}_k$ is provided by a different equation, valid from the roughness crest downwards, which takes an exponential form based on the von Karman–Pohlhausen integral method.
In this region, the velocity profile is assumed to take the form
\begin{equation} \label{exp law yang}
     \bar{u}(y)=\bar{u}_k e^{[a(y-k)/k]}.
\end{equation}

Here, $a$ is an attenuation factor that incorporates the sheltering mechanism of individual roughness elements, which can promote or reduce the ability of the flow to 'skim over' the roughness pattern, thus modulating the expected drag.
The virtual origin $d$, present in Eq. \eqref{differential log law}, is also a function of $a$, and for details on their expression and derivation we refer to the study of \citet{yang2016exponential}.

By applying Eq. \eqref{exp law yang} in our compressible database, we assume that the flow in the roughness sublayer is almost at stagnation, so density variations are small enough to consider the exponential layer still valid without compressibility corrections.
Under this assumption, an analytical solution of Eq. \eqref{differential log law} exists for $y=k$, where $\bar{\rho}_k \approx \bar{\rho}_w$, which reads
\begin{equation}\label{eq:loganalytic}
    \bar{u}_k = \frac{u_\tau}{\kappa} \ln{\frac{k-d}{z_0}},
\end{equation}
where $z_0$ is the hydrodynamic roughness length, which is a function of the flow properties and elements size, shape and arrangement, as described in \citet{yang2016exponential}. From the roughness crest upwards, $y>k$, Eq. \eqref{differential log law} is integrated numerically accounting for mean density variations (i.e. is coupled to the thermodynamic field).

The solution process for the present model is as follows. 
First, Eqs. \eqref{exp law yang} and \eqref{eq:loganalytic}, together with the geometrical constraints provided by \citet{yang2016exponential}, are used to estimate the ratio $\bar{u}_k/u_\tau$. Then, a first guess of $u_\tau$ is made, and the differential equation \eqref{differential log law} is integrated from the roughness crest using the velocity $\bar{u}_k$, up to the matching location of choice.
Here, Eq. \eqref{differential log law} is coupled to the temperature equation \eqref{eq griffin} through the density ratio provided by the \citet{van1951turbulent} transformation, which is computed through the equation of state.
After the integration process, the friction temperature can be obtained by evaluating Eq. \eqref{walz 3} at a certain location, which we choose to be the roughness crest $k$, with the expression
\begin{equation}\label{eq T_tau}
     T_\tau = \bigg(\bar{T}_k - \bar{T}_w + r_g\frac{\bar{u}_k^2}{2c_p}\bigg)\frac{\bar{u}_k}{u_\tau Pr}.
\end{equation}
Finally, heat transfer is obtained using the definition of friction temperature $q_w=T_\tau c_p \rho_w u_\tau$.
The choice of using $\bar{u}_k$ and $\bar{T}_k$ in Eq. \eqref{eq T_tau} has been made because the value $\bar{u}_k/u_\tau$ yields directly from the sheltering mechanism, being directly proportional to the turbulence spreading angle \citep{yang2016exponential}. Even though in this location the GRA does not perform the best (see Section \S \ref{sec:validation}), this approach avoids potential inaccuracies and stiffness in the velocity and temperature coupling that could arise when evaluating these quantities at other locations through integration procedures.
In summary, the heat flux predictive capabilities of the GRA have been coupled to a physics-based model for drag through the use of the compressibility transformation of \citet{van1951turbulent}. 
 \vspace{-0.1cm}
\section{\textit{A priori} results}
\begin{figure}%[t]
  %\centerline{\fbox{\vbox to 8pc{\hbox to 10pc{}}}}
  %\includegraphics[scale=.4]{sample.eps}
  \centering
	\subfigure[][]{\includegraphics[scale=0.7]{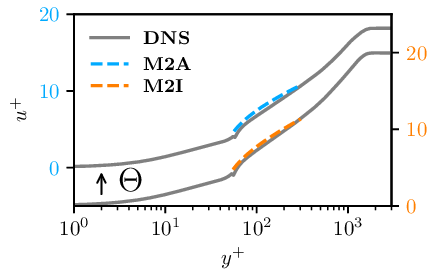}}\qquad
	\subfigure[][]{\includegraphics[scale=0.7]{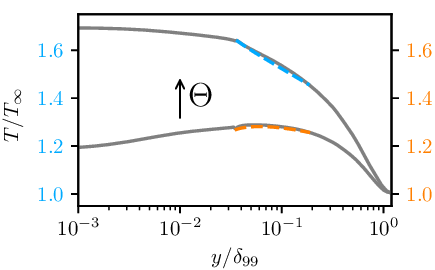}}\\
    \vspace{-10pt}
    \subfigure[][]{\includegraphics[scale=0.7]{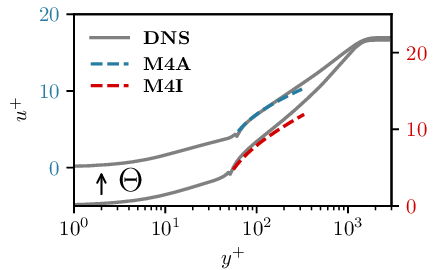}}\qquad
	\subfigure[][]{\includegraphics[scale=0.7]{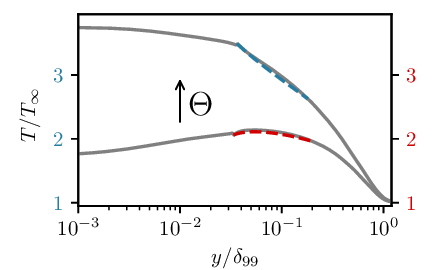}}
    \vspace{-10pt}
  \caption{Mean velocity (a,c) and temperature (b,d) profiles as function of the wall normal coordinate $y^+$ and $y/\delta_{99}$, respectively. Panels (a,b) report cases at $M_\infty=2$, while Panels (c,d) at $M_\infty=4$.
  Each figure shows two wall temperature conditions: adiabatic ($\Theta\approx1$) and cold wall ($\Theta=0.25$). For the velocity profiles, adiabatic cases (M2A and M4A) are manually shifted upwards by $\Delta u^+=5$ (left axis) in order to distinguish them from cold wall cases (right axis).
  The matching location for the model is located at $y^+=300$.}
  \label{fig a_priori_results_}
\end{figure}

In this section, the present model is tested \textit{a priori} using the DNS simulations of the present dataset, listed in Table \ref{tab DNS parameters}. In particular, the mean DNS flow field is sampled at the matching location, and related flow variables are fed to the model, which in response provides an estimate of mean velocity and temperature profiles, as well as wall fluxes $\tau_w$ and $q_w$.
Figure \ref{fig a_priori_results_} shows the model's prediction of the mean velocity and temperature profiles of the adiabatic and isothermal boundary layers over aligned cubical elements. Panels (a-b) show cases at $M_\infty =2$, M2A and M2I, while Panels (c-d) show cases at $M_\infty =2$, M4A and M4I. 
Here, the model outputs are shown from the roughness crest $y=k$, up to the matching location $y=y_m$, which is located at $y^+=300$, close to the upper limit of validity of the logarithmic law for the present dataset.
In general, there is an excellent agreement between the model outputs and the reference DNS profiles, with minor deviations only for the velocity profiles at $M_\infty=4$.
We emphasize that in these plots the values of $\bar{u}_k$ and $\bar{T}_k$ are outputs of the model, which are in excellent agreement with the predicted value by DNS data.

We then report the relative percentage error in the prediction of wall shear stress $\epsilon_{\tau_w}$, defined as $\epsilon_{\tau_w}= 100 (\tau_{w,DNS}-\tau_{w,model})/\tau_{w,DNS}$.
Cases at $M_\infty =2$ show $\epsilon_{\tau_w}=1.77 \%$ for M2A and $\epsilon_{\tau_w}=-0.69 \%$ for M2I.
At $M_\infty =4$, we obtain $\epsilon_{\tau_w}=-10.77 \%$ for M4A and $\epsilon_{\tau_w}=-20.94 \%$ for M4I.
The respective counterpart for the wall heat flux $\epsilon_{q_w}$, with an analogous definition, amounts to $\epsilon_{q_w}=5.53 \%$ for case M2I and  $\epsilon_{q_w}=-0.37 \%$ for case M4I, respectively.
For adiabatic cases, $\epsilon_{q_w}$ is undefined, but we confirm that the output heat flux from the model is close to zero, as expected.

Both cases at $M_\infty =2$ show errors below $6 \%$ in both $\tau_w$ and $q_w$, reflecting the excellent agreement observed for velocity and temperature profiles in Panels (a-b) of Figure \ref{fig a_priori_results_}.

Cases at $M_\infty=4$ show a slight increase of errors for $\tau_w$ while the prediction of $q_w$ for case M4I is still excellent.
We attribute the higher error in $\tau_w$ for this case mainly to inaccuracies introduced from the \citet{van1951turbulent} velocity transformation, which stands out in Figure \ref{up} for not being able to perfectly collapse this case onto the others.

The significance of the present results becomes particularly evident when compared with one of the state-of-the-art approaches for estimating heat transfer over rough walls, namely the correlation proposed by \citet{hill1980measurements}. When applied to the present cases, this model yields heat-flux errors of $\epsilon_{q_w}=192\%$ for case M2I and $216\%$ for case M4I. Moreover, the Hill correlation requires as inputs the rough-wall skin-friction coefficient $C_f$, the smooth-wall reference quantities $C_{f,s}$ and $C_{h,s}$, as well as the roughness Reynolds number $k^+$, whereas the present approach achieves substantially higher accuracy without relying on these additional wall-model parameters.

 \vspace{-0.3cm}
\section{Conclusions}
In this study, we leverage a novel DNS database spanning multiple Mach numbers and wall thermal conditions to assess the validity of the generalized Reynolds analogy (GRA) for compressible rough-wall flows. We show that the local GRA remains asymptotically valid in the outer layer, while the momentum–energy analogy breaks down below the roughness crest. Outer-layer velocity similarity is recovered through the transformation of \citet{van1951turbulent}, which successfully collapses both smooth- and rough-wall compressible flows onto incompressible references, enabling wall-model development within an incompressible framework. Based on these findings, we propose a new rough-wall wall model that couples GRA-based heat-transfer prediction with the drag model of \citet{yang2016exponential}, with velocity and temperature linked through density. The model shows excellent \textit{a priori} performance at $M_\infty=2$, with errors below $6\%$ for both $\tau_w$ and $q_w$, and maintains accurate heat-flux predictions at $M_\infty=4$ despite a slight degradation in drag prediction.

Future works are needed to investigate different flow conditions, as well as specific sets of roughness geometries, which may introduce stronger compressibility effects. To this end, we underline that the building blocks of the proposed model rely on physics-based assumptions, such that individual components can be modified to accommodate realistic roughness geometries and more advanced compressibility transformations, while requiring substantially fewer tunable parameters than traditional semi-empirical correlations.

\backsection[Acknowledgements]{We acknowledge that the results reported in this paper have been achieved using the EuroHPC JU Extreme Scale Access Infrastructure resource Marenostrum 5 hosted at BSC-CNS, Barcelona, Spain, under project EHPC-EXT-2023E01-034.We also acknowledge the CINECA award under the ISCRA and EuroHPC initiatives (project EUHPC\_E02\_044), for the availability of high-performance computing resources on Leonardo booster.}

\backsection[Funding]{We acknowledge financial support under the National Recovery and Resilience Plan (NRRP), Mission 4, Component 2, Investment 1.1, Call for tender No. 104 published on 2.2.2022 by the Italian Ministry of University and Research (MUR), funded by the European Union – NextGenerationEU– Project Title ADMIRE - CUP B53C24006770006 - Grant Assignment Decree No. 1401 adopted on 18/09/2024 by the Italian Ministry of Ministry of University and Research (MUR). This research received also financial support from ICSC – Centro Nazionale di Ricerca in ``High Performance Computing, Big Data and Quantum Computing'', funded by European Union – NextGenerationEU.}
\backsection[Declaration of interests]{The authors report no conflict of interest.}

\backsection[Data availability statement]{The data that support the findings of this study are available upon reasonable request.}
 \vspace{-0.2cm}

\bibliographystyle{jfm}
\bibliography{jfm}

@article{breugem2005direct,
  title={Direct numerical simulations of turbulent flow over a permeable wall using a direct and a continuum approach},
  author={Breugem, W.-P. and Boersma, B.-J.},
  journal={Phys. Fluids},
  volume={17},
  number={2},
  year={2005},
  publisher={AIP Publishing}
}

@article{duan2011direct,
  title={Direct numerical simulation of hypersonic turbulent boundary layers. {P}art 4. {E}ffect of high enthalpy},
  author={Duan, L. and Martin, M.P.},
  journal={J. Fluid Mech.},
  volume={684},
  pages={25--59},
  year={2011},
  publisher={Cambridge University Press}
}

@book{walz1969boundary,
  title={Boundary layers of flow and temperature},
  author={Walz, A.},
  year={1969},
  publisher={MIT Press}
}

@article{zhang2014generalized,
  title={A generalized {R}eynolds analogy for compressible wall-bounded turbulent flows},
  author={Zhang, Y. S. and Bi, W. T. and Hussain, F. and She, Z. S.},
  journal={J. Fluid Mech.},
  volume={739},
  pages={392--420},
  year={2014},
  publisher={Cambridge University Press}
}

@article{van1951turbulent,
  title={Turbulent boundary layer in compressible fluids},
  author={Van Driest, E. R.},
  journal={J. Aeronaut. Sci.},
  volume={18},
  number={3},
  pages={145--160},
  year={1951}
}

@article{crocco1932sulla,
  title={Sulla trasmissione del calore da una lamina piana a un fluido scorrente ad alta velocità},
  author={Crocco, L.},
  journal={L’Aerotecnica},
  volume={12},
  pages={181--197},
  year={1932}
}

@book{busemann1931gasdynamik,
  title={Gasdynamik},
  author={Busemann, A.},
  year={1931},
  publisher={Akademische Verlagsgesellschaft}
}

@book{townsend1980structure,
  title={The structure of turbulent shear flow},
  author={Townsend, A. A. R.},
  year={1980},
  publisher={Cambridge university press}
}

@article{wenzel2022influences,
  title={About the influences of compressibility, heat transfer and pressure gradients in compressible turbulent boundary layers},
  author={Wenzel, C. and Gibis, T. and Kloker, M.},
  journal={J. Fluid Mech.},
  volume={930},
  year={2022},
  pages={A1},
  publisher={Cambridge University Press}
}

@article{bernardini2023streams,
  title={STREAmS-2.0: Supersonic turbulent accelerated Navier-Stokes solver version 2.0},
  author = {Bernardini, M. and Modesti, D. and Salvadore, F. and Sathyanarayana, S. and Della Posta, G. and Pirozzoli, S.},
  journal={Comput. Phys. Commun.},
  volume={285},
  pages={108644},
  year={2023},
  publisher={Elsevier}
}

@article{Modesti_2022, title={Direct numerical simulation of supersonic turbulent flows over rough surfaces}, volume={942}, DOI={10.1017/jfm.2022.393}, journal={J. Fluid Mech. }, author={Modesti, D. and Sathyanarayana, S. and Salvadore, F. and Bernardini, M.}, year={2022}, pages={A44}}

@article{owen_heat_1963,
	title = {Heat transfer across rough surfaces},
	volume = {15},
	copyright = {https://www.cambridge.org/core/terms},
	issn = {0022-1120, 1469-7645},
	url = {https://www.cambridge.org/core/product/identifier/S0022112063000288/type/journal_article},
	doi = {10.1017/S0022112063000288},
	language = {en},
	number = {3},
	urldate = {2025-05-26},
	journal = {J. Fluid Mech.},
	author = {Owen, P. R. and Thomson, W. R.},
	month = mar,
	year = {1963},
	pages = {321--334},
}

@article{chen_compressible_1972,
	title = {Compressible {Turbulent} {Boundary}-{Layer} {Heat} {Transfer} to {Rough} {Surfaces} in {Pressure} {Gradient}},
	volume = {10},
	issn = {0001-1452, 1533-385X},
	url = {https://arc.aiaa.org/doi/10.2514/3.50166},
	doi = {10.2514/3.50166},
	language = {en},
	number = {5},
	urldate = {2025-10-16},
	journal = {AIAA J.},
	author = {Chen, K. C.},
	month = may,
	year = {1972},
	pages = {623--629},
}

@inproceedings{hill1980measurements,
  title={Measurements of surface roughness effects on the heat transfer to slender cones at Mach 10},
  author={Hill, J and Voisinet, R and Wagner, D},
  booktitle={18th Aerospace Sciences Meeting},
  pages={345},
  year={1980}
}

@article{chung2021predicting,
  title={Predicting the drag of rough surfaces},
  author={Chung, D. and Hutchins, N. and Schultz, M. P. and Flack, K. A.},
  journal={Annu. Rev. Fluid Mech.},
  volume={53},
  pages={439--471},
  year={2021},
  publisher={Annual Reviews}
}

@article{jimenez2004turbulent,
  title={Turbulent flows over rough walls},
  author={Jim{\'e}nez, J.},
  journal={Annu. Rev. Fluid Mech.},
  volume={36},
  pages={173--196},
  year={2004},
  publisher={Annual Reviews}
}

@article{cogo2023assessment,
  title={Assessment of heat transfer and Mach number effects on high-speed turbulent boundary layers},
  author={Cogo, M. and Ba{\`u}, U. and Chinappi, M. and Bernardini, M. and Picano, F.},
  journal={J. Fluid Mech.},
  volume={974},
  pages={A10},
  year={2023},
  publisher={Cambridge University Press}
}

@article{yang2016exponential,
  title={Exponential roughness layer and analytical model for turbulent boundary layer flow over rectangular-prism roughness elements},
  author={Yang, X. I. A. and Sadique, J. and Mittal, R. and Meneveau, C.},
  journal={J. Fluid Mech.},
  volume={789},
  pages={127--165},
  year={2016},
  publisher={Cambridge University Press}
}

@article{cogo2025a,
  title={Surface roughness effects on subsonic and supersonic turbulent boundary layers},
  author={Cogo, M. and Modesti, D. and Bernardini, M. and Picano, F.},
  journal={J. Fluid Mech.},
  volume={1009},
  pages={A56},
  year={2025},
  publisher={Cambridge University Press}
}

@article{cogo2025b,
  title={Development of supersonic turbulent boundary
layers over prism-shaped rough surfaces},
  author={Cogo, M. and Modesti, D. and Picano, F. and Bernardini, M.},
  journal={J. Fluid Mech.},
  volume={1025},
  pages={A21},
  year={2025},
  publisher={Cambridge University Press}
}

@article{manzoor2024estimating,
  title={Estimating mean profiles and fluxes in high-speed turbulent boundary layers using inner/outer-layer scalings},
  author={Hasan, M. A. and Larsson, J. and Pirozzoli, S. and Pecnik, R.},
  journal={AIAA J.},
  volume={62},
  number={2},
  pages={848--853},
  year={2024},
  publisher={American Institute of Aeronautics and Astronautics}
}

\end{document}